# An individual-based model of collective attention

Mehdi Moussaid[1,2], Dirk Helbing[2], Guy Theraulaz[1]

[1] Centre de Recherches sur la Cognition Animale, CNRS, Université Paul Sabatier, Toulouse, France; [2] ETH Zurich, [2] Swiss Federal Institute of Technology, Chair of Sociology, Zürich, Switzerland.

Phone: +33 5 61 55 64 41

Fax: +33 5 61 55 61 54

Email: mehdi.moussaid@gmail.com

In our modern society, people are daily confronted with an increasing amount of information of any kind. As a consequence, the attention capacities and processing abilities of individuals often saturate. People, therefore, have to select which elements of their environment they are focusing on, and which are ignored. Moreover, recent work shows that individuals are naturally attracted by what other people are interested in. This imitative behaviour gives rise to various herding phenomena, such as the spread of ideas or the outbreak of commercial trends, turning the understanding of collective attention an important issue of our society. In this article, we propose an individual-based model of collective attention. In a situation where a group of people is facing a steady flow of novel information, the model naturally reproduces the log-normal distribution of the attention each news item receives, in agreement with empirical observations. Furthermore, the model predicts that the popularity of a news item strongly depends on the number of concurrent news appearing at approximately the same moment. We confirmed this prediction by means of empirical data extracted from the website *digg.com*. This result can be interpreted from the point of view of competition between the news for the limited attention capacity of individuals. The proposed model, therefore, provides new elements to better understand the dynamics of collective attention in an information-rich world.

*Collective behaviour – attention – information flow- digg.com*

## Introduction

With the ongoing growth of mass media and communication technologies, people are every day confronted with an increasing amount of information of any kind. Pieces of information can be broadcasted through television, urban advertisements or Internet, or exchanged directly between people through emails, phone calls or personal communications.

As a side effect, the amount of information an individual is facing in its everyday life often exceeds its attention capacities[1]. Therefore, people have to *select* which elements of their environment they are focusing on, and which are ignored or checked out roughly. Interestingly, when selecting items they pay attention to, individuals are strongly influenced by other people's choices. First, because people like to share the same topics of interest as their friends, neighbors or colleagues, second, because popular novels are more relayed in the media, which increases their level of attraction and, finally, because people are naturally curious about what others are interested in[2].

The study of collective attention, therefore, is an important issue in the understanding of various herding phenomena, such as the spread of ideas[3], the dynamics of donations[4], or the outbreak of commercial trends resulting in bestsellers or blockbusters[5].

Recent work shows that some patterns of collective attention can be understood by invoquing epidemic models. For example, the temporal evolution of video views on *YouTube.com* is well described by a macroscopic model, where first viewers trigger a cascade of subsequent views over a social network[5]. Similarly, the donation rate after the Asian tsunami and various aspects of collective attention on the interactive website *digg.com* can be understood by means of similar processes[4][6].

In this contribution, we propose an individual-based model of collective attention, describing how people behave and interact when facing a steady flow on novel information. We first describe in more details the phenomena under study and highlight some empirical patterns of collective attention, as observed on the World Wide Web. Then follows a description of the model that we suggest to explain the above patterns. Finally, we present first simulation results and see how well they match the above observations and what we can learn from them.

## Patterns of collective attention

In the age of the World Wide Web, scientists have gained access to unprecedented volumes of data on human activities[5-8]. The interactive website *digg.com*, for example, is an interesting source of data to observe peoples' behaviors in situations where they have to select relevant information from a very large pool of items. The website allows its users to submit contents found elsewhere on the web. Each submitted story appears on the website and is dynamically assigned to an explicit measure of popularity, which motivates people to read it or not. If a user likes a story, he can add a "digg" to it. The total number of diggs a story has received is displayed next to it, and provides an indicator of its level of popularity. If a submission receives enough diggs within a certain time period, it eventually jumps to the 'popular' section and becomes accessible by the users more easily.

We have tracked the evolution of the digg rate for 1137 popular stories submitted from July 2 to July 12, 2007. The digg rate $\gamma$ is defined as the number of diggs that a story has received during a certain period of time $\delta t$ (here we choose $\delta t =10$ minutes). This enabled us to track the evolution of a story's popularity over time (Fig. 1a). The average curve displays a typical popularity pattern, characterized by a sudden burst of attention shortly after the story comes out, followed by a slow relaxation to low popularity (Figure 1a). Alternatively, one can observe how often a specific keyword has been searched on *Google* over time[1] [5]. As an illustration, the search pattern for the keyword '*tsunami*' during the Asian catastrophe in 2004 is displayed in the inset of figure 1a. We observe a very similar dynamics, suggesting that this pattern is not specific to *digg.com* but holds a universal feature regarding phenomena of collective attention.

.

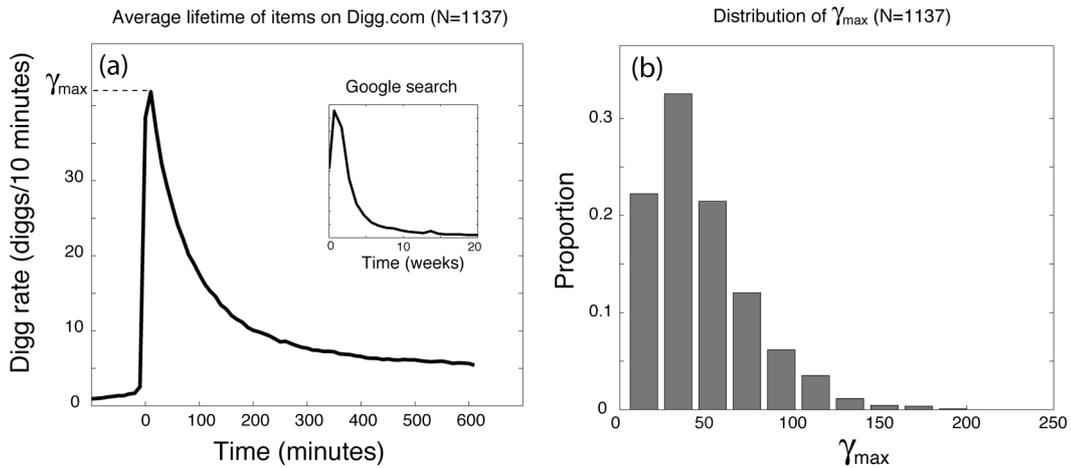

**Figure 1**. (a) Average amount of attention given to stories on digg.com over time. Time t=0 is the time at which the story appeared in the popular section. The inset indicates the amount of search queries for the keyword 'tsunami' on google over time, after the Asian tsunami happened in December 2004. The value $\gamma_{max}$ indicates the maximum popularity level that the news item has reached. (b) The distribution of the values of $\gamma_{max}$ for the same dataset.

We then focused on the differentiation between the items, that is on how much the maximum level of popularity differed between the stories. Figure 1b shows the distribution of the maximum digg rate $\gamma_{max}$ each story reached. It turns out that most stories are barely paid attention to, while a few of them attracted the attention of a large number of users. A Kolmogorov-Smirnov normality test of $\log(\gamma_{max})$ with mean 3.66 and standard deviation 0.64 yields a P-value of 0.0413, suggesting that $\gamma_{max}$ follows a log-normal distribution. In the following, we consider this statistical feature as a signature of the phenomenon and use it to test and validate our model, which we describe in the following section.

---
[1] '*Google trends*': http://www.google.com/trends

## Model description

In the field of human psychology and neurobiology, it is well-known that the brain has a limited rate of information processing. In particular, one person can either *select* a particular item of the environment for detailed analysis and completely ignore other items, or it can *share* the attention among several targets while missing the details[10]. Accordingly, we suggest that each agent $j$ of our model has a limited attention capacity $C_j$ that can be shared over many items. The agent can, for example, pay 100% of its attention to a single event, or 70% to a first one and 10% to three others, and so forth. The value of $C_j \geq 0$ is initially set to 1 for all agents.

The environment in which agents interact is defined as a collection of N items, each item $i$ being characterized by its age $A_i$ and popularity $P_i$. The age of each item naturally increases in time, while the popularity is the amount of attention that the item receives from the agents. Once an item reaches a certain maximum age $A_{max}$, it is removed.

Given this framework, we rely on earlier empirical observations to define three behavioral rules that describe the agents' behavior.

**Rule 1.** The first behavioral rule defines how the actors of the model are attracted by the news items. It describes the fact that, in real life, the probability of an uninformed individual to hear about a novelty increases with its popularity, as demonstrated by Stanley Milgram during early experiments[2].

To formalize this idea, we assign to each item $i$ a weight $W_i(t)$ that describes its attractiveness**.** The greater the weight of an item, the higher its probability to be chosen. $W_i(t)$ is defined according to Milgram's observations:

$$W_i(t) = 1 - e^{-k_1 P_i(t)} + \varepsilon$$

where the parameter $k_1$ describes how strong is the effect of increasing popularity and $\varepsilon$ is a fluctuation term defined as a uniformly distributed random value selected in the range 0 to $\varepsilon_0$.

At the end, each agent $j$ chooses an item, with a probability

$$p_i(t) = \frac{W_i(t)}{\sum_{n=1}^{N} W_n(t)}.$$

**Rule 2.** Once the agent has selected an item, it spends a fraction $\beta_i$ of its attention capacity to it. A recent study based on the website *digg.com* shows that, as an item becomes older, it attracts less attention[6]. Accordingly, we define the fraction $\beta_i$ as a function of the age $A_i$ of a news item by:

$$\beta_i(t) = e^{-k_2 A_i(t)}$$

The amount $\beta_i$ is then subtracted from the attention capacity $C_j$ of the agent and added to the popularity $P_i$ of the item. If the corresponding $\beta_i$ would exceed the remaining capacity $C_j$ of the agent, the value $C_j$ is taken instead (so that the attention capacity of the agents can never fall below zero).

**Rule 3.** Finally, we consider that agents continuously recover a small amount $\alpha$ of attention they previously gave to each item. For all items $i$ currently attended by agent $j$, the amount $\alpha$ is deduced from popularity $P_i$ and goes back to its attention capacity $C_j$. As for the previous rule, if $\alpha$ exceeds the amount of attention agent $j$ is giving to item $i$, this latter value is taken instead, so that the total amount of attention in the system remains constant.

## Numerical investigation of the model

We have then conducted a set of simulations based on the three rules describes above. The population was set to n=100 agents, each having an initial attention capacity $C_j = 1$. At each time step, a news item is added with probability $p_{in}$ describing the inflow of information.

As expected, novel items are eventually discovered by agents at random a short time after their occurrence and attract a variable amount of attention from the agents. Because of their increasing age, the popularity level of older items decreases, and they vanish some time later. This dynamics yields an average curve of popularity that is similar to the observed ones (see Fig. 2a). Moreover, we have defined the variable $P_{max}$ as the maximum attention that an item has attracted during its lifetime (i.e. the peak value of the attention burst). The distribution of $P_{max}$ follows, in fact, a log-normal distribution (Pvalue=0.23). This result is in good agreement with observations described in section 2 and reflects the fact that most items are barely considered by the agents, while a few of them reach a high level of popularity (Fig. 2b).

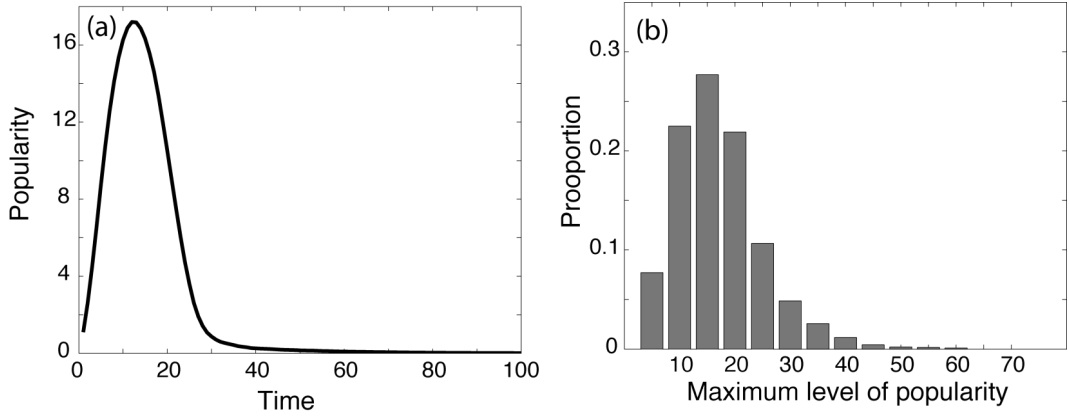

**Figure 2.** Simulation results. (a) Average popularity dynamics of the item. (b) Distribution of the maximum level of popularity. The parameters used in our simulation are as follows: n=100 agents; $p_{in}$=0.25; $\varepsilon_0$=0.2; $\alpha$=0.15; $A_{max}$=100; $k_1$=0.1; $k_2$=0.05

To investigate on the effects of competition between items, we have measured the cumulated amount of attention $P_{cumul}$ each item attracted during its lifetime as a function of the news inflow $p_{in}$. The model predicts a decay of the average popularity with an increasing flow of information (figure 3a). Further on, we have investigated whether this competition effect is also visible on digg.com. For this, we have determined the correlation between the average final number of diggs of stories and the number of concurrent stories posted within the previous or following 30 minutes. For all 37316 popular stories posted between July 2007 and June 2008, we found that the average number of diggs a story reached decreases with the number of concurrent stories (Fig. 3b). An exponential decay fits the observed dataset well (R-square=0.84), in accordance with the model prediction. Interestingly, this dynamics has been observed on digg.com regardless of the news content, topic or relevance, suggesting that the quality of the item would often have a minor influence on the collective dynamics as compared to the timing of its release.

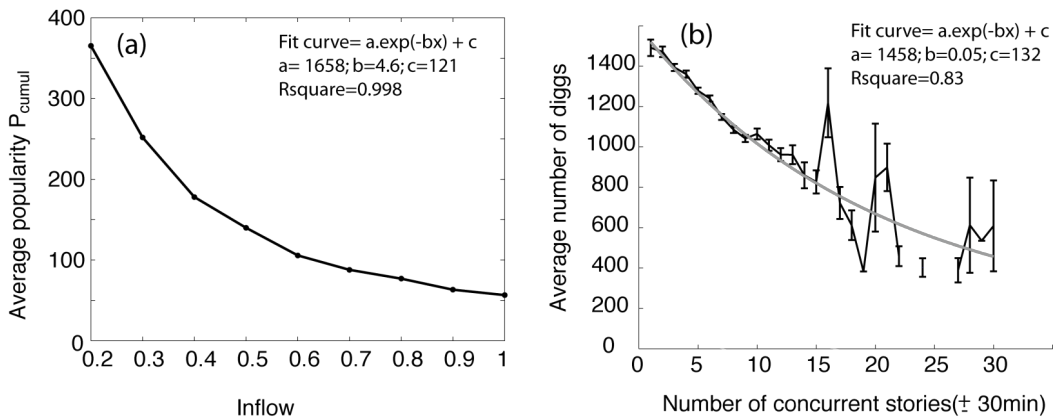

**Figure 3.** Effect of competition between concurrent news. (a) The simulation model predicts the average amount of attention for each news item decreases with increasing arrival of news. (b) The average final number of diggs for 37316 stories on digg.com as a function of the number of concurrent stories posted within the same period of time. Error bars indicate the standard error of the mean value.

## Conclusion

We have presented a simple individual-based model of collective attention, which ingredients are mainly based on empirical observations. Our simulation results display several encouraging similarities with the dynamics observed on the website *digg.com*. The model predicted that the expected popularity of a news item strongly depends on the number of items appearing approximately at the same time. This could be confirmed by data from digg.com. We expect this as an effect of competition for a limited attention capacity. Our observations may be compared with situations where several commercial products (e.g. movies or books) are released at almost the same time [11].